\PassOptionsToPackage{draft}{hyperref}
\documentclass{ANStopical}
\usepackage{xstring}
\usepackage{booktabs}
\usepackage{mdframed}
\newif\ifincludeoptional
\includeoptionalfalse 
\newmdenv[linewidth=0.8pt,linecolor=black,
  leftline=true,rightline=true,topline=true,bottomline=true,everyline=false,
  innerleftmargin=4pt,innerrightmargin=4pt,
  skipabove=6pt,skipbelow=6pt]{expandednote}
\graphicspath{{figures/}}
\newcommand{\tauint}{\tau_{\mathrm{int}}}
\newcommand{\E}{\mathbb{E}}
\newcommand{\RefBranch}{38.64264}
\newcommand{\RefAbs}{30.43135}
\newcommand{\DomainL}{200}
\newcommand{\OpticalL}{33}
\newcommand{\ScatterRatio}{0.9}
\newcommand{\NRep}{28}

\newcommand{\NInitial}{1,000,000}
\newcommand{\NGen}{2000}

\newcommand{\Moment}{2.0000045\pm0.0000080}
\newcommand{\OffspringMean}{1.0000022\pm0.0000040}

\newcommand{\CutoffShiftBranch}{0.00668}
\newcommand{\CellsError}{4.51}

\newcommand{\CosineError}{0.55}

\newcommand{\LegendreError}{1.69}

\newcommand{\CellsBranchlessError}{4.21}
\newcommand{\CellsBranchingNFError}{4.52}

\newcommand{\CosineBranchlessError}{2.56}
\newcommand{\CosineBranchingNFError}{0.55}
\newcommand{\CosineCompression}{4.6}
\newcommand{\CosineCompressionMin}{4.6}
\newcommand{\CosineCompressionMax}{4.7}
\newcommand{\LegendreBranchlessError}{8.03}
\newcommand{\LegendreBranchingNFError}{1.70}
\newcommand{\LegendreCompression}{4.7}
\newcommand{\LegendreCompressionMin}{4.4}
\newcommand{\LegendreCompressionMax}{4.8}
\newcommand{\CellResidualRange}{0.0054\text{--}0.0092}
\newcommand{\CellResidualSERange}{4.0\text{--}8.7}
\newcommand{\ProfileReplicas}{5}
\newcommand{\ProfileInitial}{100,000}
\newcommand{\ProfileGenerations}{200}
\newcommand{\TransportMin}{12.60}
\newcommand{\TransportMax}{16.10}
\newcommand{\CellsCost}{2.101\pm0.015}
\newcommand{\CellsStorage}{20.48}
\newcommand{\CosineCost}{1.792\pm0.025}
\newcommand{\CosineStorage}{2.56}
\newcommand{\LegendreCost}{3.006\pm0.023}

\newcommand{\CosineSolveShare}{23}
\newcommand{\LegendreSolveShare}{55}
\newcommand{\LegendreSolveRatioMin}{2.1}
\newcommand{\LegendreSolveRatioMax}{2.5}

\title{An Analytic Two-Dimensional Benchmark for\\
Integrated Autocorrelation Times in Branching Monte Carlo\\
Part II: Operator Estimation from Functional Expansion Tallies}
\addAuthor[jlmiao@psu.edu]{Jilang Miao}{1}
\addAuthor{Guillaume L. Giudicelli}{2}
\addAuthor{William Reed Kendrick}{3}
\addAffiliation{1}{Department of Nuclear Engineering,
Pennsylvania State University, University Park, PA 16802}
\addAffiliation{2}{Idaho National Laboratory, Idaho Falls, ID 83415}
\addAffiliation{3}{Department of Nuclear Science and Engineering,
Massachusetts Institute of Technology, Cambridge, MA 02139}
\shortTitle{Branching Monte Carlo Benchmark: FET Operators}
\authorHead{J. Miao et al.}
\Abstract{Functional expansion tallies can represent transport operators for
Monte Carlo correlation prediction without first constructing a fine cell
transition matrix. We tally Galerkin operators directly from
birth-to-absorption histories, retaining the sampled departure measure
through an empirical mass matrix. Operator moments and an estimated
offspring second moment give branching autocorrelations and integrated
autocorrelation times without an eigendecomposition. The analytic benchmark
from Part I supplies continuum and noise-free finite-model references.
At the compared sizes, cosine and Legendre models have smaller IACT errors
and require less operator storage than the cell model, and the branching
reweighting reduces their representation error by a further factor near five
while leaving the cell error unchanged. The measured runtime benefit is basis
dependent. Profiling separates shared transport from the
additional costs of operator assembly and prediction.}

\begin{document}

\section{Introduction}

Correlations between Monte Carlo source generations affect tally variance
and statistical convergence~\cite{Miao2016}. Predicting them supports
uncertainty estimation and the design of source updates for variance
reduction~\cite{Miao2020}. Multitype branching process (MBP) models track
source transport and multiplication in discrete phase-space
regions~\cite{Miao2018}, but their dense matrices require storage quadratic
in the number of regions.

Functional expansion tallies (FET) estimate spatial distributions through
basis coefficients. For smooth distributions, they can improve on histogram
approximations~\cite{Griesheimer2006}; statistical criteria can guide
expansion order~\cite{Wang2021}. Replacing cell indicators by a smaller
functional basis provides the model order reduction (MOR) investigated here.

Physically, the operator estimated here maps the birth site of a source neutron
to the site where its history ends.
Repeated application describes how the fission source spreads through the domain
from one generation to the next,
and its slowly decaying modes are the spatial patterns that persist across generations
and correlate successive tallies.
In practice, an ordinary run records the birth and absorption positions of each history
and accumulates products of basis functions evaluated at those two positions,
together with the offspring counts, as additional tallies.
After the run, a small linear system built from these tallies is solved repeatedly
to propagate the tally through successive generations.
The output is the predicted autocorrelation function of the tally
and its integrated autocorrelation time,
which give the variance inflation of the tally average.

Part I supplies analytic transport spectra and branching covariances for
assessment. Because such spectra are unavailable in general problems, we
estimate finite operators from histories. We extend the earlier
scattering-chain comparison~\cite{Miao2026FET} to absorption and critical
branching. Cell and Galerkin operators are tallied directly from the same
histories without a fine-cell matrix as an intermediate. An empirical mass matrix and an
estimated offspring second moment give
autocorrelation functions (ACFs) and integrated autocorrelation times
(IACTs) through operator moments. Noise-free references separate
representation error from fitted-model differences. We compare accuracy
and storage, and separate shared transport from additional pair-product
assembly in the runtime profiles.

\section{Benchmark and Prediction Targets}

Neutrons undergo planar isotropic scattering in $\Omega=[-L,L]^2$ with reflective
boundaries, exponential flights of rate $\Sigma_t$, and scattering probability
$c_s<1$. Let $K$ map an isotropic departure site to the next collision
site, and let $A$ map a birth site to its eventual absorption site.
The first argument is the departure point and the operator acts on functions:
$(Af)(\mathbf x)=\int_\Omega A(\mathbf x,\mathbf y)f(\mathbf y)
\,d\mathbf y$.
For the scattering chain, each collision is followed by a new isotropic
flight. For the absorption/rebirth chain, each absorbed particle is replaced
by one particle at its absorption site.

In the branching process, an absorbed parent produces a nonnegative integer
number $\xi_a$ of children at that site. The offspring law is independent
of position and transport, with $\E\xi_a=1$ and finite raw second moment
$m_2=\E\xi_a^2$. Children are transported independently conditional on
their birth sites. Capture corresponds to $\xi_a=0$; the derivation does
not require a fixed nonzero offspring count. The source bank is not
normalized to a fixed size. Its tally is the fraction of source particles
in $I=[-\alpha L,\alpha L]^2$, defined while the bank is nonempty.

Both $K$ and $A$ are reversible for the uniform
stationary density $\pi=1/(4L^2)$. Their complete orthonormal eigenfunctions
$e_{np}$ are products of normalized shifted cosines with wavenumbers
$k_n=n\pi/(2L)$. Inner products are
$\langle f,g\rangle_\pi=\int_\Omega fg\,\pi\,d\mathbf x$.
The absorption eigenvalues are
\begin{equation}
 \Lambda_{np}=\frac{\Sigma_a}
 {\sqrt{\Sigma_t^2+k_n^2+k_p^2}-\Sigma_s},\qquad
 \Sigma_s=c_s\Sigma_t,\quad\Sigma_a=(1-c_s)\Sigma_t.
 \label{eq:spectrum}
\end{equation}
The collision eigenvalues are
$\mu_{np}=\Sigma_t/\sqrt{\Sigma_t^2+k_n^2+k_p^2}$, and
$\Lambda_{np}=(1-c_s)\mu_{np}/(1-c_s\mu_{np})$.
Thus scattering and absorption use the same eigenfunctions but different
lag units: collisions and absorption events, respectively. Criticality
makes $A$ the mean reproduction operator for the branching process as well.
The nonconstant eigenvalues lie below one. For the centered tally
$h=\mathbf1_I-\pi_I$, where $\pi_I=\alpha^2$, and
$W_\kappa=\langle h,e_\kappa\rangle_\pi^2$, Part I derives the branching
weights
\begin{equation}
 W_\kappa^{\mathrm{br}}=
 W_\kappa\frac{m_2-\Lambda_\kappa^2}{1-\Lambda_\kappa^2}
 =W_\kappa\left[1+\frac{m_2-1}{1-\Lambda_\kappa^2}\right].
 \label{eq:weights}
\end{equation}
These weights describe equilibrated nonconstant source modes after replacing
the fluctuating tally denominator by its mean population. This is a
large-population approximation on a finite observation window.

For either nonbranching chain, the continuum ACF is
$\rho(d)=\sum_{\kappa\ne0}W_\kappa\theta_\kappa^d/
\sum_{\kappa\ne0}W_\kappa$, where $\theta_\kappa$ is the corresponding
collision or absorption eigenvalue. For branching, replace $W_\kappa$ by
$W_\kappa^{\mathrm{br}}$ and use $\theta_\kappa=\Lambda_\kappa$.
In each case $\tauint=1+2\sum_{d\geq1}\rho(d)$, provided the sum
converges. These are the continuum targets. The following sections obtain
finite-model predictions without using the known eigensystem.

\section{Direct Tallies of the Operator}
In a general transport problem no spectral solution is available, so the operator must be estimated from the histories themselves. The fitted models are therefore tallied from Monte Carlo histories, in either cells or a functional basis, and the spectral solution of this benchmark serves only to assess them.

We use $M$
for the total number of cells and $r$ for the total number of basis functions,
including the constant. For tensor-product models,
$M=M_{\mathrm{side}}^2$ and $r=r_b^2$. The absorption operator is tallied
from birth-to-absorption pairs. The same construction estimates $K$ when
the pairs are successive collision sites; branching additionally requires
the offspring moment.

\subsection{Cell Transition Counts}

For each simulated parent history, record its birth position $\mathbf x_j$
and absorption position $\mathbf y_j$. Capture histories are included,
and a fission history is counted once rather than once per child. Let
$C_{ab}$ count pairs from cell $a$ to cell $b$, and
$v_a=\sum_b C_{ab}$. With $S=\sum_a v_a$ histories,
\begin{equation}
 \widehat P_{ab}=\frac{C_{ab}}{v_a},\qquad
 \widehat\pi_a=\frac{v_a}{S},\qquad
 h_a=\mathbf1_I(a)-\widehat\pi_I,\qquad
 \widehat\pi_I=\sum_{a\subset I}\widehat\pi_a.
 \label{eq:cells}
\end{equation}
The tally boundary aligns with the grids considered here.
Rows of unvisited cells are set to zero. The departure frequencies
$\widehat\pi$ estimate the sampling measure. 

\subsection{Functional Bases and the Mass Matrix}

Let $b(\mathbf x)$ contain $r$ linearly independent basis functions, with
$b_0=1$. Define the Galerkin matrices under a departure measure $\psi$ by
\begin{equation}
 \mathsf M_{kl}=\langle b_k,b_l\rangle_\psi,\qquad
 G_{kl}=\langle b_k,A b_l\rangle_\psi.
 \label{eq:populationmatrices}
\end{equation}
Testing an approximate eigenfunction $b^Tv$ against every basis function
gives $Gv=\lambda\mathsf Mv$. This generalized problem depends on the
measure used for both matrices. 

The same history pairs used in Eq.~\eqref{eq:cells} directly accumulate
\begin{align}
 \widehat G&=\frac1S\sum_{j=1}^S b(\mathbf x_j)b(\mathbf y_j)^T,&
 \widehat{\mathsf M}&=\frac1S\sum_{j=1}^S b(\mathbf x_j)b(\mathbf x_j)^T,
 \label{eq:pairtallies}\\
 \widehat b_h&=\frac1S\sum_{j=1}^S b(\mathbf x_j)
       [\mathbf1_I(\mathbf x_j)-\widehat\pi_I],&
 \widehat\pi_I&=\frac1S\sum_{j=1}^S\mathbf1_I(\mathbf x_j).
 \label{eq:tallyvector}
\end{align}
Here $\widehat{\mathsf M}$ records how the sampled birth sites are spread over the basis patterns,
$\widehat G$ records how birth sites in one pattern feed absorption sites in another,
and $\widehat{\mathsf M}^{-1}\widehat G$ is the generation-to-generation transfer between patterns.
These are product tallies at known birth and absorption sites. They do not
require the analytic kernel, a mesh-based precursor matrix, or its
eigenvectors. Source particles share ancestry, so contributions to these
matrices are not independent observations. Sampling errors below are
estimated from independent replicas of the entire calculation.

With cell indicators $b_a=\mathbf1_a$,
$\widehat{\mathsf M}=\operatorname{diag}(v/S)$ and
$\widehat G=C/S$. Consequently,
$\widehat{\mathsf M}^{-1}\widehat G=\widehat P$ on visited cells.
The mass matrix is exactly the departure-count normalization that converts
counts to transition probabilities. This places the cell model in the same projection framework as the other bases.

The cosine models use the constant function and $\sqrt2\cos(m\pi x/L)$.
The polynomial models use normalized Legendre functions
$\sqrt{2m+1}P_m(x/L)$. Both use tensor products, with
$0\leq m<r_b$ in each direction. Cosines span the even transport sector
and provide an invariant-subspace reference for the symmetric tally.
Legendre functions are not transport eigenfunctions. They provide a test of
direct operator estimation without prescribing the eigenbasis. Their retained
dimension includes odd functions even though the noise-free symmetric tally
has zero projection onto odd sectors.

\section{Correlation Prediction From Operator Moments}

A finite operator could be diagonalized and its modes substituted into a
spectral formula. 
However, a full dense eigensolve has cubic cost and,
for sampled matrices that are nonsymmetric in the empirical inner product, 
yields complex eigenvalues and mode coefficients. 
Instead, we propagate the tally vector and form real scalar
moments, avoiding the full eigensolve.

\subsection{The Moment Sequence}

Write $U=K$ for collision steps or $U=A$ for absorption steps, and let
$\theta_\kappa$ denote the corresponding eigenvalue, $\mu_\kappa$ or
$\Lambda_\kappa$, following the notation of Part I. Conditional
expectation for either stationary chain gives
\begin{equation}
 c_d=\E[h(\mathbf x_0)h(\mathbf x_d)]
 =\langle h,U^dh\rangle_\pi
 =\sum_{\kappa\ne0}W_\kappa\theta_\kappa^d.
 \label{eq:moments}
\end{equation}
Centering $h=\mathbf1_I-\pi_I$ removes the constant eigenfunction. For
cells the estimate substitues $\widehat P$ and $\widehat\pi$ into Eq.~\eqref{eq:moments}, giving 
$c_d^{\mathrm{cell}}=(\widehat\pi\odot h)^T\widehat P^dh$,
where $\odot$ denotes componentwise multiplication and the cell vector
$h$ is centered with $\widehat\pi_I$ as in Eq.~\eqref{eq:cells}.
The directly tallied nonbranching prediction is therefore
\begin{equation}
 \widehat\rho(d)=\frac{c_d^{\mathrm{model}}}{c_0^{\mathrm{model}}},\qquad
 \widehat\tauint=1+2\frac{\sum_{d\geq1}c_d^{\mathrm{model}}}
                              {c_0^{\mathrm{model}}}.
 \label{eq:nonbranchingprediction}
\end{equation}
Here the superscript model identifies a cell or Galerkin approximation.
For scattering, this supplies the
derivation of the operator-based prediction considered in the earlier
study~\cite{Miao2026FET}. For absorption, only the histories used to tally
the operator and the physical meaning of a lag change.

For a Galerkin representation under the measure $\psi$, let
$h_r(\mathbf x)=b(\mathbf x)^Tq$ be the projection of the centered tally.
The following relations apply to the matrices in
Eq.~\eqref{eq:populationmatrices}; their fitted counterparts use the
empirical matrices and tally vector in Eqs.~\eqref{eq:pairtallies}
and~\eqref{eq:tallyvector}. Requiring
$\langle b_k,h_r\rangle_\psi=\langle b_k,h\rangle_\psi$
for every $k$ gives $\mathsf Mq=b_h$. Applying $A$ and projecting again
gives coefficient vector $Tq$, with $T=\mathsf M^{-1}G$. Repetition yields
\begin{equation}
 c_d^{(r)}=q^T\mathsf M T^dq=b_h^TT^dq,\qquad
 c_0^{(r)}=b_h^Tq.
 \label{eq:galmoments}
\end{equation}
This propagates the projected operator at each step, which coincides with
projecting $A^d$ once only in an invariant subspace. Its normalization
$c_0^{(r)}$ is the variance of the projected tally; using the full indicator
variance $\pi_I(1-\pi_I)$ instead would define a different approximation.
For scattering, replace $A$ by $K$ in the definition of $G$; the mass
matrix, projection, and propagation algebra are unchanged.

The implementation iterates $v_{d+1}=\widehat Pv_d$ from $v_0=h$ for cells
and, for Galerkin, factors $\widehat{\mathsf M}$ once and solves
$\widehat{\mathsf M}v_{d+1}=\widehat Gv_d$ from $v_0=q$; a dot product
with the left tally vector gives $c_d$, and no eigenvectors are needed.

Moment propagation gives the ACF for both scattering and
absorption/rebirth. If only IACT is required, however, their single lag
sum in Eq.~\eqref{eq:nonbranchingprediction} can be evaluated directly
by a constrained linear solve, without computing each lag separately.
Branching introduces an offspring-dependent covariance correction, which
the following subsection evaluates through four moment sums.

We first derive the direct solve for a stationary nonbranching model.
For scattering, $G_{kl}=\langle b_k,Kb_l\rangle_\psi$ and lag counts
collisions, as in the earlier study~\cite{Miao2026FET}. For
absorption/rebirth, $G_{kl}=\langle b_k,Ab_l\rangle_\psi$ and lag counts
absorption events. These are different transport matrices, but both
operators preserve constants and lead to the constrained formulation
below. Define
$s=\sum_{d\geq1}T^dq$ on the centered, decaying subspace. Subtracting the
same series shifted by one power gives $(I-T)s=Tq$, or
\begin{equation}
 (\mathsf M-G)s=Gq.
 \label{eq:resolvent}
\end{equation}
Let $e_0$ represent the constant basis function. Since the operator
preserves constants, $Ge_0=\mathsf Me_0$, and $\mathsf M-G$ is singular.
Imposing zero mean, $e_0^T\mathsf Ms=0$, gives the constrained
Karush--Kuhn--Tucker (KKT) system
\begin{equation}
 \begin{bmatrix}\mathsf M-G&\mathsf Me_0\\e_0^T\mathsf M&0\end{bmatrix}
 \begin{bmatrix}s\\\eta\end{bmatrix}
 =\begin{bmatrix}Gq\\0\end{bmatrix},\qquad
 \tauint^{(r)}=1+2\frac{b_h^Ts}{b_h^Tq}.
 \label{eq:kkt}
\end{equation}
With empirical matrices this is a constrained estimator; the measured
departure distribution need not be exactly stationary.

\subsection{Branching Reweighting}

We now use $U=A$, so the moments describe birth-to-absorption transport.
The position pairs determine the mean spatial operator, but not the
fluctuations in offspring count. The latter enter through $m_2$.
The offspring moment and the mean reproduction rate are tallied as
\begin{equation}
 \widehat m_2=\frac1S\sum_{j=1}^S\xi_{a,j}^2,\qquad
 \widehat\nu=\frac1S\sum_{j=1}^S\xi_{a,j}.
 \label{eq:offspringestimate}
\end{equation}
Capture events contribute zero to the sums but remain in $S$. The
critical model uses mean reproduction one; $\widehat\nu$ is reported as a check of criticality.

To convert the branching spectral formula into a matrix calculation,
first interchange the lag sum in IACT with the mode sum in the ACF.
Summing the geometric lag series gives a factor
$\Lambda_\kappa/(1-\Lambda_\kappa)$ for each mode.
Equation~\eqref{eq:weights} then separates its covariance weight into a
branchless contribution and a multiplication correction. Expanding the
resulting rational functions as power series converts each mode sum back
into weighted sums of $c_n=\sum_\kappa W_\kappa\Lambda_\kappa^n$.
These moments are the inner products already computed by matrix
propagation, so no eigenvalues are required in the fitted calculation.
All mode sums below exclude the constant mode.

The four required sums are
\begin{align}
 S_0&=\sum_\kappa W_\kappa=c_0,&
 S_1&=\sum_\kappa\frac{W_\kappa\Lambda_\kappa}{1-\Lambda_\kappa}
       =\sum_{n\geq1}c_n,\label{eq:sumsone}\\
 T_0&=\sum_\kappa\frac{W_\kappa}{1-\Lambda_\kappa^2}
       =\sum_{k\geq0}c_{2k},&
 T_1&=\sum_\kappa\frac{W_\kappa\Lambda_\kappa}
                {(1-\Lambda_\kappa)(1-\Lambda_\kappa^2)}.
 \label{eq:sumstwo}
\end{align}
The last expression is also a moment sum,
$T_1=\sum_{n\geq1}\lceil n/2\rceil c_n$.
The prediction is
\begin{equation}
 \widehat\tauint^{\mathrm{br}}
 =1+2\frac{S_1+(\widehat m_2-1)T_1}{S_0+(\widehat m_2-1)T_0}.
 \label{eq:branchingtau}
\end{equation}
All four sums are accumulated as running totals during the single moment iteration that generates $c_n$.
The branchless diagnostic is
$1+2S_1/S_0$, and the same sequence gives the branching ACF as
\begin{equation}
 \widehat\rho^{\mathrm{br}}(d)=
 \frac{c_d+(\widehat m_2-1)\sum_{k\geq0}c_{d+2k}}
 {c_0+(\widehat m_2-1)\sum_{k\geq0}c_{2k}}.
 \label{eq:branchingacf}
\end{equation}
The fitted prediction uses only the tallied matrices and the estimated
$\widehat m_2$; the analytic solution serves only for assessment.

The series require decay of the contributing nonconstant modes for convergence. The code
tests $n|c_n|<10^{-10}|c_0|$ for 20 consecutive lags, with a cap of 20000
lags, and reports failed convergence instead of a prediction. The factor
$n$ accounts for the increasing coefficient in $T_1$.

Finite sampling makes $\widehat G$ nonsymmetric, so its eigenvalues can be
complex; the moment expressions remain real matrix functionals whenever the
series converge and need no diagonalization. Applied to a sampled operator
they are a plug-in approximation to this reversible benchmark. 

\section{Noise-Free References}

For assessment we distinguish three quantities. The continuum reference
uses the analytic spectrum and the specified offspring law. The fitted
prediction uses Monte Carlo matrices and the estimated offspring moment. The
noise-free finite-model reference uses the same retained representation as
the fitted model, but constructs its matrices by deterministic integration
and uses the specified $m_2$. Each comparison uses the same physical target:
$m_2=1$ for absorption/rebirth, the specified offspring law for branching.

Noise-free references are computed for each representation independently
of the Monte Carlo pairs. For cells,
\begin{equation}
 P_{ab}^{\mathrm{nf}}=\frac{1}{\pi_a}
  \int_a\pi(\mathbf x)\int_b A(\mathbf x,\mathbf y)
           \,d\mathbf y\,d\mathbf x.
 \label{eq:cellreference}
\end{equation}
The cell integrals are evaluated using the analytic eigenfunctions and a
truncated kernel expansion. The spectral cutoff introduces numerical
reference error independently of the cell representation.

For Galerkin, deterministic integration supplies $G$, $\mathsf M$, and
$b_h$, followed by a symmetric generalized eigenproblem. Cosine entries
are diagonal in the noise-free invariant subspace. Legendre entries use
quadrature of their overlaps with the transport modes. Even modes suffice for
the centered symmetric tally; the Monte Carlo operator also carries odd
blocks, which couple only through sampling noise.

Applying the same branching reweighting of Eq.~\eqref{eq:weights} to each finite-model spectrum separates
representation error from the difference between fitted and noise-free
models. The latter difference can include matrix sampling, offspring-moment
estimation, finite-sample nonlinear bias, and numerical reference error, so
it is not purely statistical. Deterministic references are calculated once
per configuration.

\section{Numerical Assessment}
\subsection{Simulation and Reference Parameters}

The operator estimates use the same benchmark as the direct Monte Carlo
validation in Part I, with $L=\DomainL\,\mathrm{cm}$,
$\Sigma_tL=\OpticalL$, $c_s=\ScatterRatio$, and $\alpha=1/2$.
The optical size and central tally region are unchanged from the earlier
scattering benchmark~\cite{Miao2026FET}. Offspring counts
$\xi_a\in\{0,2\}$ have equal probabilities, giving $\E\xi_a=1$ and
$m_2=2$. This introduces nonzero offspring variance while preserving
criticality. Evaluating the continuum mode sums through even-sector
spatial cutoff 500 gives IACTs of \RefAbs{} for absorption/rebirth and
\RefBranch{} for branching, with lag summed to infinity.
The production finite-model references use
even-sector kernel cutoff 64 for cells, and cutoff 80 with 512-point
Gauss--Legendre quadrature for Legendre.

The accuracy ensemble contains \NRep{} independent branching replicas,
each initialized with \NInitial{} uniform source particles and followed
for \NGen{} generations. All parent histories contribute to the matrices
and offspring moments, including those before the source-ACF burn-in.
The source intensity is uniform in expectation from initialization.
Matrices for all requested orders are taken from the largest assembled
basis in each replica. This reuse is appropriate for an accuracy sweep;
the separate cost study below assembles only its specified order.

\subsection{Fitted Predictions}

Each branching prediction uses the offspring second moment estimated from
its own replica, retaining its dependence on the operator tallies rather
than substituting the prescribed moment. The ensemble estimates of both
offspring moments agree with their specified values within one standard
error. All production IACT predictions are finite.
Table~\ref{tab:accuracy} compares noise-free and fitted predictions for
both absorption/rebirth and branching. The absorption/rebirth columns use
the same birth-to-absorption matrices tallied in the branching replicas and
omit the multiplication correction, so they assess the mean operator
separately from the branching reweighting.

\begin{table}[!htb]
 \centering
 \caption{Noise-free and fitted IACTs. Absorption/rebirth uses $m_2=1$;
 branching uses $m_2=2$ for references and replica-estimated $\widehat m_2$
 for fits. Uncertainties are standard errors across \NRep{} replicas.}
 \label{tab:accuracy}
 \begin{tabular}{lrrrr}
\toprule
& \multicolumn{2}{c}{Absorption/rebirth} & \multicolumn{2}{c}{Branching} \\
\cmidrule(lr){2-3}\cmidrule(lr){4-5}
Model & Noise-free & Fitted & Noise-free & Fitted\\
\midrule
Continuum & 30.4314 & {} & 38.6426 & {} \\
\midrule
Cells, $M=1600$ & 29.1489 & $29.1533\pm0.0013$ & 36.8956 & $36.9010\pm0.0013$ \\
Cosine, $r=400$ & 31.2090 & $31.2083\pm0.0013$ & 38.8562 & $38.8553\pm0.0014$ \\
Legendre, $r=400$ & 32.8740 & $32.8731\pm0.0013$ & 39.2979 & $39.2970\pm0.0014$ \\
\bottomrule
\end{tabular}

\end{table}

\begin{figure}[!htb]
 \centering
 \includegraphics[width=\textwidth]{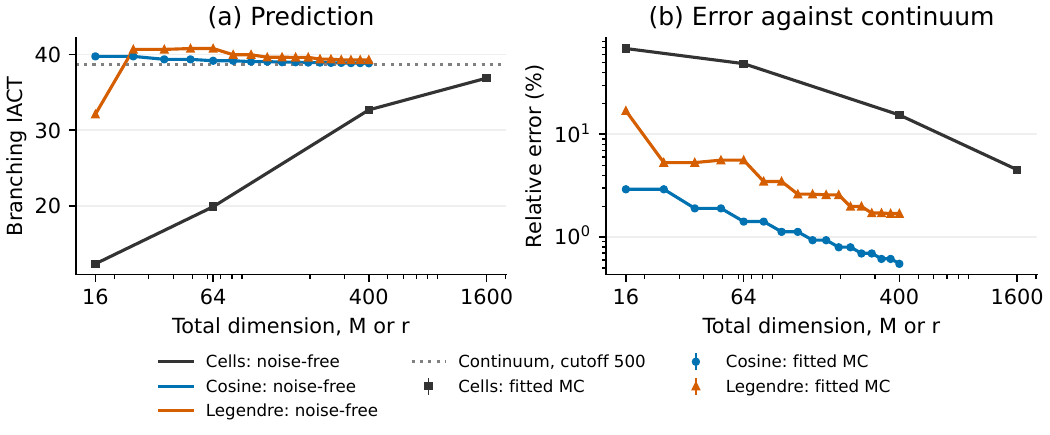}
 \caption{Branching IACT and relative error against the cutoff-500 continuum
 reference. Lines denote noise-free models and markers fitted means; fitted
 standard errors in the left panel are smaller than most markers.}
 \label{fig:accuracy}
\end{figure}

Figure~\ref{fig:accuracy} shows the full dimension dependence. At $r=400$,
cosine and Legendre errors against the stated continuum reference are
\CosineError\% and \LegendreError\%, compared with \CellsError\% for
$M=1600$. Cosine benefits from retaining exact eigenfunctions of the
benchmark; Legendre tests a basis without this advantage. The Legendre
sequence is not monotone at small orders, so a small error at an isolated
order is not sufficient evidence of convergence. Plateaus also occur when
new functions have zero noise-free projection by symmetry.

The fitted cosine and Legendre values at $r=400$ differ from their
noise-free counterparts by less than one standard error. Cell values exceed
their references by $\CellResidualRange$, corresponding to
$\CellResidualSERange$ standard errors. These small residuals are much
smaller than the cell representation errors but cannot be described as
agreement within sampling uncertainty. Further kernel-cutoff and estimator
bias checks are needed to identify their origin.

The branching reweighting changes the representation error of the two kinds
of model differently. Against their respective continuum values, the
noise-free cosine error falls from \CosineBranchlessError\% for
absorption/rebirth to \CosineBranchingNFError\% for branching at $r=400$,
a factor of \CosineCompression; the Legendre error falls from
\LegendreBranchlessError\% to \LegendreBranchingNFError\%, a factor of
\LegendreCompression. The ratio is nearly constant over $r\geq100$ for both
bases (\CosineCompressionMin{} to \CosineCompressionMax{} and
\LegendreCompressionMin{} to \LegendreCompressionMax), so it belongs to
smooth-basis projection rather than to the cosine functions being
eigenfunctions. The cell error changes only from \CellsBranchlessError\% to
\CellsBranchingNFError\%. Equation~\eqref{eq:weights} explains the
difference: the factor $1+(m_2-1)/(1-\Lambda_\kappa^2)$ weights slowly
decaying modes most. A smooth basis represents these accurately and confines
its error to the fast modes it truncates, where the weight is smallest,
whereas cell averaging distorts every mode, including the slow ones.

There is also uncertainty in treating a finite spatial sum as a continuum
reference. The independent cutoff check in Part I changes the branching
reference by \CutoffShiftBranch{} when the cutoff increases from 500 to
2000. We retain the saved cutoff-500 convention for all error curves and
percentages here, so the accuracy comparison is meaningful at the percent
scale rather than to every displayed digit.

\section{Computational Cost and Storage}

Transport and branching are performed on the GPU. Birth and absorption
coordinates are processed in chunks; basis arrays supply matrix products
for $G$ and $\mathsf M$. Cell counting in this campaign uses a dense CPU
histogram after downloading coordinates. Galerkin matrices are assembled
on the GPU, downloaded, and normalized; mass factorization and moment
iteration are performed on the CPU.

For $S$ histories, a cell count update has constant work per pair, apart
from the dense histogram initialization and reduction. Dense Galerkin
assembly costs $O(Sr^2)$ and also requires basis evaluation. Reduction in
matrix dimension does not remove this per-history work. The stored cell
transition matrix costs $O(M^2)$, whereas $G$ and $\mathsf M$ cost
$O(r^2)$. With $D$ moment iterations, the prediction costs
$O(DM^2)$ for cells and $O(r^3+Dr^2)$ for Galerkin, including mass
factorization.

The dedicated profiling campaign uses one NVIDIA A100-PCIE-40GB and eight
allocated Intel Xeon Gold 6248R CPU cores. Each configuration has
\ProfileReplicas{} replicas with \ProfileInitial{} initial particles and
\ProfileGenerations{} generations. The actual history counts fluctuate
with bank size, but the same five bank histories are reused across all
configurations. The basis sweep uses $r_b=10,20,40,60,80$; the cell sweep
uses $M_{\mathrm{side}}=4,8,20,40,80$. Each basis profile assembles only
the specified order, not all smaller orders.

\begin{table}[!htb]
 \centering
 \caption{Profiling times per replica in seconds (mean $\pm$ standard error),
 excluding transport and branching. Size denotes total cells or basis functions.}
 \label{tab:cost}
 \begin{tabular}{lrrrr}
\toprule
Model & Size & Assembly/tally & Transfer & Solve\\
\midrule
Cells & 16 & $1.108\pm0.012$ & $0.14492\pm0.00159$ & $0.00126\pm0.00002$ \\
 & 64 & $1.101\pm0.012$ & $0.15329\pm0.00150$ & $0.00308\pm0.00008$ \\
 & 400 & $1.084\pm0.011$ & $0.14447\pm0.00107$ & $0.059\pm0.001$ \\
 & 1600 & $1.732\pm0.013$ & $0.16113\pm0.00157$ & $0.208\pm0.003$ \\
 & 6400 & $22.445\pm0.056$ & $0.20976\pm0.00231$ & $6.099\pm0.113$ \\
\midrule
Cosine & 100 & $0.410\pm0.101$ & $0.000214\pm0.000005$ & $0.022\pm0.004$ \\
 & 400 & $1.618\pm0.025$ & $0.00127\pm0.00012$ & $0.173\pm0.001$ \\
 & 1600 & $14.412\pm0.198$ & $0.01443\pm0.00102$ & $2.124\pm0.014$ \\
 & 3600 & $77.625\pm0.995$ & $0.08965\pm0.00033$ & $14.812\pm0.013$ \\
 & 6400 & $237.728\pm2.593$ & $0.28199\pm0.00161$ & $54.522\pm0.121$ \\
\midrule
Legendre & 100 & $0.719\pm0.003$ & $0.000180\pm0.000004$ & $0.045\pm0.001$ \\
 & 400 & $2.587\pm0.016$ & $0.000896\pm0.000016$ & $0.418\pm0.008$ \\
 & 1600 & $17.516\pm0.243$ & $0.01186\pm0.00136$ & $5.235\pm0.117$ \\
 & 3600 & $81.078\pm0.870$ & $0.08664\pm0.00021$ & $36.624\pm0.755$ \\
 & 6400 & $245.259\pm2.517$ & $0.27846\pm0.00078$ & $134.137\pm2.957$ \\
\bottomrule
\end{tabular}

\end{table}

In Table~\ref{tab:cost}, the cell solve includes count normalization,
Galerkin transfer includes download and normalization, and Galerkin solve
includes mass factorization and moment iteration; ACF extraction, submatrix
selection, and reference construction are excluded. Each run includes both
bases and a placeholder model, so a method's phase sum is its accounted cost
rather than a standalone runtime.

At $M=1600$ the accounted cell cost, including coordinate transfer, is
$\CellsCost$ seconds; at $r=400$ it is $\CosineCost$ seconds for cosine and
$\LegendreCost$ seconds for Legendre. Cosine therefore improves both the
error and the accounted cost at these sizes, whereas Legendre improves the
error at a higher cost. The comparison is at the selected sizes rather than
at matched error.

The Legendre solve takes \LegendreSolveRatioMin{} to
\LegendreSolveRatioMax{} times as long as the cosine solve at equal $r$.
One moment iteration costs the same for both bases at equal $r$, so the
fitted Legendre operator needed correspondingly more iterations: its slowest
contributing mode decays more slowly. At $r=6400$ the solve is
\CosineSolveShare\% of assembly for cosine and \LegendreSolveShare\% for
Legendre, no longer a minor cost.

Transport timings range from \TransportMin{} to \TransportMax{} seconds
across configurations despite identical bank histories, so the phase
comparisons do not constitute a controlled end-to-end speedup measurement.
Phase-sum uncertainties are formed within each replica.

\begin{figure}[!htb]
 \centering
 \includegraphics[width=\textwidth]{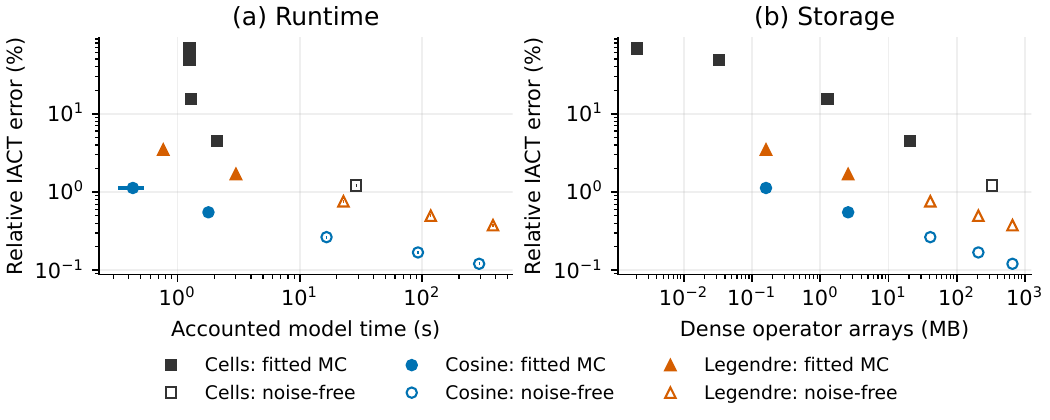}
 \caption{IACT error against the cutoff-500 continuum reference versus
 accounted time and operator storage. Filled markers use production fitted
 means; open markers use noise-free models at larger sizes without production
 fits. All timings measure MC assembly, transfer, and prediction, excluding
 shared transport, branching, and reference construction. Horizontal bars
 show one standard error across profiling replicas.}
 \label{fig:cost}
\end{figure}

Figure~\ref{fig:cost} includes all profiled sizes; the open markers use
noise-free errors because no production fit exists at those sizes. A float64
$M=1600$ cell matrix occupies \CellsStorage{} MB, against \CosineStorage{} MB
for the two $r=400$ Galerkin matrices, an eightfold reduction in operator
storage. These are operator-array sizes; counts, factors, and basis chunks
require additional memory.

\section{Conclusions}

Direct pair tallies construct cell and Galerkin operators for correlation
prediction without first tallying a fine-mesh matrix and projecting it. For the reversible critical
benchmark, operator moments and an estimated offspring second moment give
the branching IACT without an eigendecomposition. At the compared sizes both
functional bases improve accuracy and reduce operator storage, and cosine
also reduces the accounted runtime. The branching reweighting lowers the
representation error of the smooth bases by a factor near
\CosineCompression{} while leaving the cell error unchanged, because the
error of a smooth basis lies in the fast modes that branching weights least.
Pair-product assembly remains additional work, and the branching solve grows
with dimension. Numerical-reference sensitivity and small cell residuals
limit claims at finer precision. Legendre shows what a basis chosen without
the eigenfunctions achieves; a data-driven basis from the same pair tallies,
for example by proper orthogonal decomposition, could close that gap and is
left for future work.
The same pair tally over consecutive generations,
with cell indicators and the fission sites of the children as endpoints,
is the construction behind the fission matrix;
a response-matrix tally over region boundaries would follow the same pattern.
Whether the Galerkin form keeps its storage and accuracy advantages there is an open question.
The numerical example is a single homogeneous reflective square
with a fixed, position-independent offspring law and a central tally,
which is a favorable setting for a smooth basis.
The approach must be benchmarked further on leaky systems before general use.
The
predicted correlations can inform source-update design and other
variance-reduction methods; extension to heterogeneous systems requires
additional covariance modeling.

\bibliographystyle{ANStopical}
\bibliography{references}
\end{document}